\documentclass[authoryear,12pt,3p]{jowarticle}

\usepackage[english]{babel}
\usepackage[utf8]{inputenc}
\usepackage{amsmath}
\usepackage{graphicx}
\usepackage{setspace}
\usepackage[colorinlistoftodos]{todonotes}
\usepackage{natbib}

\makeatletter
\renewcommand\section{\@startsection{section}{1}{\z@}{-3.25ex plus -1ex minus -.2ex}{1.5ex plus .2ex}{\normalsize\bf}}
\renewcommand\subsection{\@startsection{subsection}{2}{\z@}{-3.25ex plus -1ex minus -.2ex}{1.5ex plus .2ex}{\normalsize\bf}}
\renewcommand\subsubsection{\@startsection{subsubsection}{3}{\z@}{-3.25ex plus -1ex minus -.2ex}{1.5ex plus .2ex}{\normalsize\bf}}

\usepackage{enumerate}
\usepackage{graphicx}
\usepackage[round]{natbib}
\usepackage{amssymb,amsmath, amscd, amsthm, mathrsfs}

\usepackage{ bbold }
\theoremstyle{definition}

\theoremstyle{remark}

\numberwithin{equation}{section}

\makeatother

\begin{document}
\begin{frontmatter}
\title{On Representational Redundancy, Surplus Structure, and the Hole Argument}

\author{Clara Bradley}
\address{Department of Philosophy \\ University of Bristol}
\author{James Owen Weatherall}\ead{weatherj@uci.edu} 
\address{Department of Logic and Philosophy of Science \\ University of California, Irvine}

\date{ }

\begin{abstract}
We address a recent proposal concerning `surplus structure' due to Nguyen et al. [ `Why Surplus Structure is Not Superfluous.' \emph{Br. J. Phi. Sci} Forthcoming.]  We argue that the sense of `surplus structure' captured by their formal criterion is importantly different from---and in a sense, opposite to---another sense of `surplus structure' used by philosophers.  We argue that minimizing structure  in one sense is generally incompatible with minimizing structure in the other sense.  We then show how these distinctions bear on Nguyen et al.'s arguments about Yang-Mills theory and on the hole argument.\end{abstract}
\end{frontmatter}
\doublespacing
\section{Introduction}\label{sec:intro}

There are several interrelated themes that arise in contemporary discussions of Einstein's `hole argument'.  One theme is largely historical: it is now widely recognized that the hole argument played a significant role in Einstein's thinking as he developed general relativity during the period from 1913 to 1915 \citep{Norton1984,Stachel}.  A second theme is essentially metaphysical.  On the classic treatment by \citet{Earman+Norton}, for instance, the argument shows that a certain kind of \emph{substantivalist}---namely, one who considers spacetime `points' to have a special ontological status, independent of or prior to the events that occur or field values that obtain there---is committed to a certain kind of indeterminism. To avoid this dismal conclusion, they argue, one must endorse a doctrine known as `Leibniz equivalence', which is meant to be a hallmark of \emph{relationism} (and thus a rejection of substantivalism).

A third theme, though rarely disentangled from metaphysical questions related to subtantivalism and relationism, is arguably of greater importance to physics.  This theme concerns whether the hole argument reveals an infelicity in the standard formalism of general relativity, in the form of \emph{surplus structure} or \emph{gauge freedom}.\footnote{For a discussion of the role of this theme in Earman's thinking on the hole argument during the 1970s and 1980s, see \citet{WeatherallStein} and references therein.}  The idea here is that the manifold substantivalist is committed to some structure---roughly, `spacetime points', though care is needed in interpreting this assertion---that the hole argument reveals is not only unnecessary for physics, but which also has undesirable consequences in the form of indeterminism.  Manifold substantivalism, meanwhile, is often taken to be suggested, or perhaps even implied, by the standard formalism of relativity. That is, the standard formalism apparently invokes or, on a natural reading, attributes to the world, precisely the surplus structure that the hole argument exposes.  The moral is then taken to be that one either needs to adopt an alternative understanding of this standard formalism---say, by adopting what is sometimes called `sophisticated substantivalism'---or else move to a different formalism altogether that excises this surplus structure.\footnote{Earman, for instance, proposed moving to Einstein algebras \citep{GerochEA} as a suitably `relationist' alternative to standard formulations of general relativity \citep{EarmanPD,Earman1986,Earman1989,EarmanWEST,Rynasiewicz1992,Bain,Rosenstock+etal}. Similar issues are at stake when, for instance, \citet[p. 31]{RovelliDisappearance} argues that the manifold is `a gauge artifact' in general relativity or \citet[p. 5]{SmolinThreeRoads} argues that there are no points in physical spacetime.  We take these arguments to assume, often implicitly, that something like manifold substantivalism is the `default' interpretation of the standard formalism, and to avoid that interpretation, one needs a formalism with a different, weaker metaphysics as its `default' interpretation. (Other authors, such as \citet{Friedman} and \citet{Field}, offer more direct arguments for positions similar to manifold substantivalism on the basis of the standard formalism of general relativity.)}  

We say this third theme is of greater importance to physics than the others because there is a connection between the search for alternative formulations of general relativity that avoid this `gauge freedom' and some approaches to quantum gravity.  Briefly, in constructing a quantum theory one generally wishes to identify (and quantize) only those degrees of freedom with physical significance.  Hence, if the standard formalism of general relativity implicitly includes surplus structure, a first step towards developing future theories might be to develop a new theory with less structure.  It is in connection with this third theme, then, that the hole argument has been of lingering significance in the development of a quantum theory of gravity.

In a pair of recent papers, \citet{WeatherallHoleArg,WeatherallUG} has argued against the view that the hole argument reveals that the standard formalism of general relativity has surplus structure. To the contrary, he argues, on a certain precise understanding of `surplus structure', general relativity should not be taken to have surplus structure at all.\footnote{Weatherall uses the expression `excess structure'; nothing turns on the difference between `excess' and `surplus' here.}  \citet{Nguyen+etal} have replied by questioning whether the notion of `surplus structure' that Weatherall proposes accurately captures what physicists have in mind when they argue that some physical theories exhibit such structure.\footnote{\citet{Nguyen+etal} focus on Yang-Mills theory, but if Weatherall's argument fails there, it will fail in general relativity too; indeed, \citet{WeatherallUG,WeatherallFBYMGR} has argued that, at least in this connection, Yang-Mills theory and general relativity are strongly analogous, and neither has surplus structure.}  Instead, they suggest, the relevant notion of `surplus structure' is one on which a theory exhibits `representational redundancy,' in the sense that a single situation can be represented in many equally good ways.\footnote{Apparently bolstering their case, Prop. 2 of \citet{WeatherallUG} is false as stated \citep{WeatherallErratum}, leading to the surprising conclusion that, on Weatherall's account, theories that are often taken to have `surplus structure' actually have \emph{less} structure than ones that are said \emph{not} to have surplus structure (as opposed to being equivalent, as Weatherall originally claimed).}  

Nguyen et al. go on to argue that what they characterize as `surplus structure' is not necessarily superfluous, in the sense of being freely eliminable.  We strongly agree with the substance of this moral, but think that their arguments are better characterized as establishing a somewhat different thesis than what they appear to state.  The reason is that they present their argument as if it is in conflict with another view, which is that one should minimize the structure of one's theories.\footnote{See, for instance, the Abstract and Introduction of their paper---and, indeed, the title, which makes sense only insofar as one might have initially thought surplus structure were superfluous.  We emphasize this point because an anonymous referee suggests that \citet{Nguyen+etal} may not have intended to reject the maxim that one should always minimize structure, but we think the plain meaning of their texts suggests otherwise.}  But we do not think there is any conflict, because the sense of `surplus structure' that they consider---that is, `representational redundancy'---is importantly different from what philosophers have generally thought of as surplus structure in a theory.\footnote{To be sure, we are not in the business of policing language: Nguyen et al. are clear and precise about what they mean by `surplus structure', and we think that, on their understanding of the expression, their argument is compelling and insightful. Our point, rather, is to clearly distinguish two different, nearly opposite, meanings of an expression both of which seem to be in use in the literature, and to emphasize that showing that surplus structure in one sense is not eliminable does not imply that surplus structure in other, very much distinct, senses is also not eliminable.}  In fact, their notion generally pulls in the opposite direction, in the sense that a theory admitting representational redundancy has \emph{less} structure than one without that redundancy.  This suggests that exhibiting `representational redundancy' should not be taken as a theoretical vice---at least not on grounds of structural parsimony.  To the contrary, we will argue, the sorts of considerations that have led philosophers to wish to excise surplus structure from theories should motivate one to \emph{increase} representational redundancy.


Our goal in the present paper is to defend the perspective just stated, and then argue that distinguishing representational redundancy from surplus structure provides insight into the third strand of literature on the hole argument described above.  In particular, we will argue, general relativity does not have surplus structure.  But it does, in several senses, admit of representational redundancy.  Keeping these separate helps clarify what the hole argument accomplishes: ultimately, we will argue, the hole argument is best seen as an argument \emph{against} the recommendation to eliminate representational redundancy from general relativity rather than an argument \emph{for} the claim that general relativity, as standardly presented, has surplus structure.  

The remainder of the paper will be structured as follows.  We begin by reviewing the arguments of \citet{Nguyen+etal} regarding `surplus structure' and `representational redundancy'.  We will then argue that in simple and intuitive examples, the precise notion of `surplus structure' that they propose should be associated with a theory having \emph{less} structure, not more.  We will then disambiguate the sense of `representational redundancy' captured by Nguyen et al.'s precise criterion from two other intuitive senses of `representational redundancy', one of which, we argue, does correspond to surplus structure.  We then bring this machinery to bear on the arguments Nguyen et al. provide concerning Yang-Mills theory, offering a different perspective on what their argument accomplishes.  Finally, we return to the hole argument in light of the forgoing discussion.  We conclude with some brief remarks about what we take the paper to have done.

\section{The Argument}\label{sec:NTW}

In this section, we review the proposal for understanding `surplus structure' given by \citet{Nguyen+etal}, and discuss how it differs from the arguments in \citet{WeatherallUG}. Since both use category theory to represent physical theories in similar ways, we will first describe the shared framework used by both approaches.\footnote{These ideas were introduced to philosophy of science by \citet{WeatherallTheoreticalEquiv} as a means of comparing different physical theories, following a suggestion by \citet{Halvorson}---though similar ideas have long been used in mathematics. A review of applications of this approach---termed `Theories as Categories of Models' by \citet{Rosenstock}---is given in \citep{WeatherallCategories}. For background on category theory, see \citep{MacLane}; for a gentler introduction, see \citep{Leinster}.}

In this framework, physical theories are represented as categories, whose objects consist of models of the theory, and arrows between the objects represent relations between the models. For the purpose at hand, we consider categories whose arrows are isomorphisms of the models, since isomorphisms are transformations that preserve structure and thus preserve representational capacity.\footnote{\citet{BarrettSS} and \citet{Rosenstock} both give reasons why, in some applications, it is important to consider more than just isomorphisms; for present purposes, little turns on whether one considers categories with a broader notion of arrow, as long as one does so consistently across all theories under discussion.} This is motivated by the idea that the structure of a model captures its representational content, and so isomorphic models are able to represent the same physical situations. 

Relations between theories are described by functors between the categories representing those theories.  These functors may be classified by what they `forget', using a scheme developed by \citet{Baezetal}. To understand the classification, we first need some terminology. A functor $F: \mathcal{C} \rightarrow \mathcal{D}$ is said to be \textit{full} if for every pair of objects $A,B$ of $\mathcal{C}$ the map $F:$ hom$(A,B)$ $\rightarrow$ hom$(F(A),F(B))$ induced by $F$ is surjective, where hom$(A,B)$ is the collection of arrows from $A$ to $B$. Similarly, $F$ is said to be \textit{faithful} if for every pair of objects the induced map on arrows is injective. Finally, $F$ is said to be \textit{essentially surjective} if for every object $X$ of $\mathcal{D}$, there is some object $A$ of $\mathcal{C}$ such that $F(A)$ is isomorphic to $X$.

Using this terminology, we say that a functor $F: \mathcal{C} \rightarrow \mathcal{D}$ forgets \textit{structure} if it is not full; it forgets \textit{stuff} if it is not faithful; and it forgets \textit{properties} if it is not essentially surjective. If $F$ is full, faithful and essentially surjective then it forgets \textit{nothing}.  In this case, $F$ is said to realize an \emph{equivalence of categories}.\footnote{We observe that there is an $n-$categorical perspective on this classification, where each of these three notions of `forgetting' correspond to forgetting structure at different `levels': forgetting properties means forgetting 0-structure; forgetting structure means forgetting 1-structure; forgetting stuff means forgetting 2-structure; and so on, where one extends these notions to a hierarchy of `essentially $k$-surjective' functors between $n-$categories \citep{Baez+Shulman}.  This alternative perspective may make it seem as if all of these notions of `forgeting' correspond to different kinds of `structure' that may be forgotten.  But what is important to emphasize is that, as we discuss below, it is 1-structure that most naturally corresponds to what is usually meant by the structure of a mathematical object or model of a physical theory.}  

\citet{WeatherallUG} argues that a theory has surplus structure relative to another if there is a functor from the first theory to the second, represented as categories, that forgets structure while preserving empirical significance.  On the other hand, \citet{Nguyen+etal} argue that there is another notion of surplus structure, operative in the physics literature,  which they call surplus* structure.  One theory has surplus* structure relative to another if there exists a functor from the first theory to the second, represented as categories, that forgets stuff while preserving empirical significance.  In this case, they argue, what is `forgotten' are extra arrows in the categories.  They argue that a theory with surplus* structure has what they call `representational redundancy'. 

We will say much more about the sense of `representational redundancy' at issue in the next section.  But first, let us say why Nguyen et al. make this proposal.  The starting point for their argument is largely sociological.  They claim that physicists and philosophers often attribute to certain theories---particularly, Yang-Mills theory---some sort of surplus or redundant structure `over and above [their] ... equivalence classes' of representationally equivalent models \citep[p. 10]{Nguyen+etal}.  That is, one observes that there are some theories wherein models related by certain transformations (`gauge transformations') are taken to have the same representational capacities, and one observes that these theories are often discussed as having something `extra'.  Nguyen et al. balk at the idea that this `something more' arises only if one neglects the transformations realizing these equivalences (which is what Weatherall's proposal amounts to) because physicists are well aware of these transformations, and do not explicitly neglect them when they make claims about the redundancy associated with these theories. And so they want to find some other sense in which a formulation of Yang-Mills theory might be said to have `surplus' over a formulation involving only equivalence classes---one on which the transformations relating equivalent models are never neglected.  

The next step of their argument is to observe that functors that forget stuff do, in fact, forget \emph{something}: namely, ways in which models are equivalent to one another. And so, surplus* structure is a candidate for something that some formulations of a theory might have that other formulations do not have.  More, it is a candidate that meets the desideratum just stated, because, as we elaborate in section \ref{sec:YM}, it turns out that there is a formulation of Yang-Mills theory that \emph{has} surplus* structure relative to a formulation using only equivalence classes, and moreover, this formulation includes all gauge transformations as arrows.  Thus, they conclude, surplus* structure is an attractive candidate for a precise way of characterizing what some formulations of theories have `over and above' formulations invoking only equivalence classes.\footnote{We remark that, although this is fair to say, adopting this prescription for what should be meant by `surplus' does not recover common claims that gauge theories exhibit `surplus' anything---because, as \citet{Nguyen+etal} go on to argue, what would putatively be `surplus' in such cases is ineliminable.}

\section{Representational Redundancy and Surplus Structure}\label{sec:W}

To evaluate Nguyen et al.'s proposal, we will now consider the sense of `representational redundancy' associated with surplus* structure (i.e., stuff).  It will be helpful to do so in the context of a simple example.

Suppose we are given a map of the Earth.  Now fix a two dimensional (real) vector space $V$.  (It is essential that we are dealing with just a vector space---not an inner product space, a norm space, or anything else.)  Imagine we are interested in representing directions on the surface of the Earth at some fixed location---say, Irvine, California---using the vectors in $V$.  One way to proceed is to choose some non-zero vector $v\in V$ and stipulate that this vector represents `North'---that is, it points in the direction of the longitudinal line passing through Irvine, towards the northern pole of the Earth.  It makes no difference at all which vector one chooses: any non-zero vector is as good as any other.  In fact, at this stage there is nothing to distinguish one non-zero vector from any other.\footnote{One way of thinking about what we have done here is to choose a particular (partial) reference relation.  We have not added any structure to $V$; we have just made a choice of mapping from $V$ to the world.}

Now suppose we would like to also represent `East'.  We once again choose some non-zero vector---$u\in V$ this time---and stipulate its meaning.  As before, we have a lot of freedom in which vector we choose, though not quite as much freedom as in choosing `North'.  This is because we have already fixed `North', and whatever else is the case, it is essential to what we mean by `North' and `East' that they are linearly independent.  So we require that $u$ be such that $u\neq \alpha v$ for any $\alpha\in\mathbb{R}$.  Aside from this, $u$ can be any vector we like.\footnote{Observe, however, that we do not have the same freedom for choosing `South', once we have chosen `North'.  In fact, a choice for `South' is (essentially) fixed by our choice of `North': `South' must be represented by (some positive multiple of) $-v$, since it is essential to `South' that it be the opposite direction of `North'.  We can drop the parentheticals if we adopt the convention, as we will in what follows, that `directions' are all represented by vectors of the same length.  But nothing that has been said thus far forces us to do this.}  

Now that representatives of `North' and `East' have been chosen, however, there are no further choices to make for which vectors represent which directions, as long we want the relations between the vectors in $V$ to reflect the spatial relations between directions as we usually understand them.\footnote{We acknowledge that the expression `as we usually understand them' is doing a fair amount of work, here.  In particular, we have fixed a meaning for `orthogonal' in both the mathematical context and in the world, and we are insisting that whatever reference relations we adopt regarding which vectors represent which direction respect those meanings.}   There is, in particular, a unique inner product $\langle\cdot,\cdot\rangle$ on $V$, up to a constant scalar factor, with the property that $u$ and $v$ are orthogonal, i.e., $\langle u,v\rangle=0$; the scale factor can be fixed as well by requiring that $u$ and $v$ both have the same length (which we will conventionally set to 1). This inner product uniquely fixes angles between all vectors.  Similarly, we can define an orientation by taking the ordered pair $(u,v)$, which, with standard sign conventions, captures the fact that the sense of rotation from `East' to `North' is counterclockwise.  In particular, given any further direction, there is a unique (unit) vector $x$ that has the correct inner products with $u$ and with $v$ to represent that direction.  

We have thus ended up with a vector space $V$, along with a lot more: we now have an inner product, an orientation, and an ordered orthonormal basis $(u,v)$.  We got all of this by making (arbitrary) choices for $u$ and $v$.  Had we made any other choices---$u'$ and $v'$, say---we would have ended up in the same place, up to unique isomorphism.  This fact makes precise the sense in which the original choices were `arbitrary'.

We could have started in a different way.  Noting, for instance, that we have a natural notion of `angle' between directions at a point on the surface of the Earth, we might have begun with an inner product space, $(V,\langle\cdot,\cdot\rangle)$.  Had we done so, assuming we wished that inner product to represent the spatial relations between directions as we generally understand them, there would have been less freedom in how we chose `North' and `East': the vector $v$ representing `North' would have been required to be an arbitrary \emph{unit} vector, and, once `North' was chosen, $u$, representing `East', would have been required to be one of the two unit vectors orthogonal to $v$.\footnote{Again, we note that the sense of `requirement' here turns on the prior assumption that any suitable reference relation for vectors in the present context should respect the plain meaning of terms such as `orthogonal'.}   If we had started with a preferred ordered orthonormal basis $(u,v)$, we would have had fewer choices to make, still.

What we are seeing here is a certain trade-off between, on the one hand, freedom in making representational choices; and on the other, mathematical structure.  When we use a vector space to represent directions in space, we have a great deal of freedom in choosing which vectors represent which directions; once we have fixed some of these choices, however, we have considerably less freedom in how we make subsequent choices.  If we begin with a vector space in which we have already defined additional relations between the vectors---one, that is, in which we have more structure defined---we do not have as much freedom in how we make these choices.  Generally speaking, \emph{more freedom} is afforded in cases where we have \emph{less structure}, and vice versa.\footnote{Assuming fixed conventions, shared across members of the comparison class, concerning what reference relations are acceptable.}

The `freedom' we have been discussing is a kind of `representational redundancy'---though it is probably better termed `representational freedom'.  In fact, as we will presently argue, this is precisely the sense of `representational redundancy' considered by \citet{Nguyen+etal}.  The `redundancy' at issue arises because the structure $V$ can represent the directions on a map in infinitely many equally good ways.  

One way to see what sort of freedom we have in making representational choices is by studying the autormorphisms---that is, the symmetries, or the isomorphisms from an object to itself---of a mathematical structure. The reason is that isomorphisms, generically, map objects to other objects that have the same structure.  We infer from this that they have the same representational capacities.\footnote{Here we are implicitly invoking a certain ideology about mathematical representation, defended by \citet{WeatherallHoleArg}.  These ideas are explored and substantially developed by \citet{Fletcher}.}  A (non-trivial) automorphism, then, can be thought of as revealing a way in which a single mathematical structure can do its representational work in multiple ways: that is, given some way in which the structure might be used to represent some situation, we can find another way in which the same structure might be used to represent the same situation by observing how the automorphism acts.

We can see this clearly in the vector space example already discussed.  If we begin with a vector space $V$, any non-zero vector $u\in V$ is related to any other non-zero vector $u'\in V$ by an automorphism, which in this case is a bijective linear transformation from $V$ to itself.  This captures the sense in which \emph{any} vector in $V$ is equally good at representing the direction `North'.  But now consider a vector space $V$ with a preferred vector $u$ already fixed.  There are now no bijective linear maps from $V$ to itself that take $u$ to $u$ and also take any other vector to $u$; but given any two vectors $v$ and $v'$, both not equal to $\alpha u$ for any $\alpha$, there always exists a bijective linear transformation on $V$ that keep $u$ fixed and maps $v$ to $v'$.  Thus we capture the sense in which once we pick a vector to represent `North', one can still choose \emph{any} other vector, not proportional to $u$, to represent `East'.  That we have `less' freedom in this second case is captured by the fact that the linear isomorphisms that preserve $u$ are naturally understood as a (proper) subgroup of the automorphisms of $V$. Thus we see a sense in which if structure $A$ has `more' symmetries than structure $B$, in the sense of there being  a natural or implicit (or, in some cases, explicit) proper embedding of the group of automorphisms of $B$ into the group of automorphisms of $A$, then $A$ has more representational redundancy than $B$.

Conversely, there is a natural sense in which if $A$ has more symmetries than $B$, we should say that $B$ has `more structure' than $A$.\footnote{This view is defended by \citet{BarrettCM1}.}  This is because the maps that preserve all the structure of $B$ also preserve all the structure of $A$, but there are further maps that preserve the structure of $A$ but do not preserve the structure of $B$---suggesting that there is something `more' to $B$ that changes even when everything about $A$ is preserved.\footnote{This intuition can be made precise in various contexts, including the first order case, where adding further relations to a theory (for instance) reduces the number of symmetries of its models. See, for instance, \citet{BarrettSS}.}  This is precisely what happens in the vector space case already discussed: a vector space endowed with an inner product has more structure---namely, the inner product---than a bare vector space, which is reflected in the fact that it has fewer symmetries.\footnote{It is perhaps worth noting that the same intuitions play out in standard discussions of classical space-time structure: Newtonian space-time has more structure than Galilean space-time, which has more structure than Leibnizian space-time; this is all reflected by the fact that Leibnizian space-time has more automorphisms than Galilean space-time, which has more automorphisms than Newtonian space-time.  These relationships are described in somewhat more detail by \citet{BarrettSTS} and \citet{WeatherallSTG} in a way that connects directly to how we discuss them here, though they were already recognized and well understood by, for instance, \citet{SteinNST} and \citet{EarmanWEST}.}

So we see that representational redundancy runs in the opposite direction to the amount of structure that a mathematical object has: more structure means less redundancy, and vice versa.  These relationships are naturally captured in the language of category theory as described in the previous section, and indeed, it is precisely these ideas that are meant to be captured by the comparisons of `structure', `stuff', and `properties' given by the classification of forgetful functors already described.  In particular, forgetting `stuff' corresponds to removing representational redundancy in the sense we have been discussing; whereas forgetting structure corresponds to removing structure.

To see how this works, consider the simplest of the examples above.  Define, for instance, a category \textbf{Vect}$_2$ whose objects are two dimensional vector spaces and whose arrows are linear transformations; and a category \textbf{OBVect}$_2$ of two dimensional vector spaces with fixed, ordered basis, with basis preserving maps as arrows.  There is a natural functor from \textbf{OBVect}$_2$ to \textbf{Vect}$_2$ that takes every vector space with basis to it underlying vector space, and arrows of \textbf{OBVect}$_2$ to their underlying linear transformations.  This functor is faithful and essentially surjective, but not full.  Thus, it forgets (only) structure.  We can also go in the opposite direction, trivially: choose any object $C$ of \textbf{OBVect}$_2$, and map all objects of \textbf{Vect}$_2$ to $C$, and all arrows to $1_C$.\footnote{This sort of `opposite direction' functor can be complicated to define (it generally, as here, involves the axiom of choice), and it does not always exist---this is why we have limited attention to a highly simplified case, to avoid complicated constructions that obscure the basic conceptual point.  In a sense, this is the core of Nguyen et al.'s argument, as we discuss in section \ref{sec:YM}.} This functor is essentially surjective and full, but it is not faithful.  And so this functor forgets (only) stuff.

There are two morals to draw from this example, which, we claim, are generic, at least among the sorts of structures used in physics.  The first is that forgetting structure and forgetting stuff, on this formal account, really do pull in opposite directions, as promised.  This is precisely because one of them involves a certain map failing to be surjective, and the other involves a certain map failing to be injective.  The second moral is that the direction in which we `forget structure' on this formal account corresponds to the direction in which we drop structure in the intuitive sense described above; whereas the direction in which we `forget stuff' corresponds to the one in which we remove representational redundancy.  It is in this sense that we claim the formal machinery recovers the more intuitive claims made above.

\section{Kinds of Representational Redundancy}\label{sec:Ladyman}

We have just argued that representational redundancy, of the sort Nguyen et al. associate with surplus* structure, has an almost inverse relationship with structure in another, intuitive sense of the term.  Increasing structure generally reduces representational redundancy; and conversely, increasing redundancy means eliminating structure.  But this claim, taken out of context, invites misunderstanding.  As we hope we have made clear, the expressions `representational redundancy' and `surplus* structure' used above have precise meanings, proposed by Nguyen et al.; we have taken on these meanings in our discussions thus far.  But there are other senses in which one might use the expression `representational redundancy'.  We now will introduce two other possible senses of `representational redundancy' and discuss how they relate to the foregoing.

To see the first of these, consider again the example discussed in the previous section.  As we remarked then, given a two dimensional vector space, any choice of two (linearly independent) vectors to represent North and East is (uniquely) isomorphic to any other choice.  One might be tempted to say that this isomorphism indicates a certain kind of representational redundancy in the vector space with ordered basis, since after all, what we see is that there are many two dimensional vector spaces with ordered bases that provide equally good representations of the cardinal directions.   But one has to be careful, because this sort of representational redundancy is not associated with surplus* structure---or surplus structure, as proposed by Weatherall.  The reason is that two categories differing only with regard to `how many' isomorphic objects are in each isomorphism class are, in general, categorically equivalent, and so one would not expect empirical-content-preserving functors between such categories to forget \emph{anything}. 

Indeed, category theory aside, this sort of representational redundancy will \emph{always} be present, at least for any theory formulated in modern mathematics with a set theoretic semantics.  This is because given a model of any theory, one can always generate new models of that theory by either applying some permutation to the domain of the model or else by choosing some other, equinumerous set, and fixing a bijection to that set.  That a theory has representational redundancy in this sense is uninteresting, at least from the perspective of how much structure the models of a theory have.  

What the existence of these isomorphic copies \emph{does} indicate is that some underlying structure---in this case, the vector space---has surplus* structure, since it is the freedom associated with choosing which vectors represent North and East that gives rise to the different, but isomorphic, vector spaces with bases. To preview what will come later, this distinction will be relevant to the hole argument, since the hole argument concerns the fact that there are distinct but isomorphic models of GR. As in the case of isomorphic representations of cardinal directions, this is not a source of surplus* structure, but we will argue that it arises because of the surplus* structure of an underlying structure, namely bare manifolds.

We now turn to a third possible notion of representational redundancy.  Consider the following example (due to James Ladyman).  One wishes to model a collection of colourless gas particles using interacting billiard balls. Now consider adding colour to the billiard balls. This adds structure to the theory, namely a `colour structure’. However, this also seems to add representational redundancy because we can use different colours to represent the same collection of gas particles: it does not matter what choice we make because we suppose that the gas particles we are trying to model do not actually have colour.\footnote{If the gas particles really did have colour, then one would want the models to be non-isomorphic. But then there would be no representational redundancy because one would think that there was in fact a correct colour attribution.} Therefore, this is an example where representational redundancy seems to correspond to surplus structure.

However, the kind of representational redundancy used in this case is different from what \citet{Nguyen+etal} discuss.  This is because the models of the gas particles related by a change in colour are not isomorphic. In other words, if the models were isomorphic, then nothing corresponding to colour would have been added to the theory because there would be no way to distinguish models with different colours.  Models differing only in colour structure would be equivalent according to the theory.  The representational redundancy comes from the fact that the non-isomorphic models---those related by a change in colour---represent observationally equivalent states of affairs, and therefore any colour can be used to represent the same situation. 

This sort of representational redundancy in fact corresponds to the notion of surplus structure proposed by Weatherall.  If one were to introduce categories of models of these two theories---one with histories of colourless billiard balls as objects and some suitable choice of arrows; and the other with histories of coloured billiard balls as objects, with arrows that, in addition to preserving whatever the arrows of the first category do, also preserve colour---then one would expect there to be a functor from the category of the theory with the colour structure to the one without that colour structure that preserves empirical significance and forgets structure (not stuff).  It is simple to see why: the arrows of the theory \emph{with} colour structure need to preserve colour structure, in addition to whatever is preserved by the arrows of the other theory.  And so one would expect a functor that preserved empirical significance to fail to be full.

What this discussion highlights is that ambiguity can arise in the use of the term `representational redundancy’.  To summarize, we have the following three, distinct and not necessarily mutually exclusive, senses of `representational redundancy'.  It can refer to:
\begin{enumerate}
\item (Surplus* structure / stuff) Situations in which a single model / mathematical structure can represent a given situation in many equally good ways; such cases are generally signaled by symmetries (automorphisms) of the models, and correspond to `surplus stuff' in the discussion above.
\item (Set theoretic semantics) Situations in which distinct but isomorphic models / mathematical structures can represent a given situation in many equally good ways; such cases are pervasive in modern applied mathematics using set theoretic semantics, and arise when there is surplus* structure of some underlying structure (including, for instance, an underlying set).
\item (Surplus structure) Situations in which distinct and \emph{non-isomorphic} models / mathematical structure can represent a given situation equally well; such cases are generally signaled by distinctions between mathematical structures that do not appear to have any physical or empirical significance, and correspond to `surplus structure' in the discussion above.
\end{enumerate}

The first of these is the sense used by \citet{Nguyen+etal}, and the third is the sense highlighted by Ladyman's example and used by Weatherall.  The crucial difference between the third sense of representational redunancy and the other two is that in the first two, the models that can play the same representational roles are equivalent according to the theory, whereas in the third, they are not equivalent: the models are distinguished from one another (in the Ladyman example, by the presence of the colour structure) even though, \emph{ex fiat}, there is no corresponding physical difference in the systems that they represent. In what follows, when ambiguity may arise, we will endeavor to refer to this list to specify the sense of `representational redundancy' at issue.

\section{Yang-Mills Theory Revisited}\label{sec:YM}

With this conceptual machinery in hand, we now return to Nguyen et al.'s arguments concerning Yang-Mills theory.  For the sake of simplicity, and following others in the literature, they focus on the case of electromagnetism, which is a (Abelian) Yang-Mills theory with structure group $U(1)$.\footnote{We remark that, although this is hardly a slight against Nguyen et al, it is not at all clear that the plausible positions in the non-Abelian case look very much like those in the $U(1)$ case.  (See, for instance, \citep{Healey}, \citep{WeatherallFBYMGR}, and \citep{Gilton} for discussions of some of the ways in which non-Abelian Yang-Mills theory resists interpretations that seem natural in electromagnetism---among which is the fact that field strength [curvature] is not a gauge-invariant quantity in non-Abelian theories.)} They make two basic arguments that are relevant to the issues now under discussion.  The first argument considers various ways of representing Yang-Mills fields on a contractible manifold $M$.  The second argument considers what happens when we relax the assumption that $M$ is contractible.\footnote{In fact, they go somewhat further than this, and make a proposal concerning how to think of the spaces of possible field configurations over all manifolds $M$ at once.  They conclude that to treat this problem adequately, one should move from thinking about theories as categories of models to thinking of theories as functors---in this case, as a functor from a category of manifolds to a category of groupoids.  This proposal has many virtues, but it does not bear directly on the issues we discuss here.}  

We begin with their first argument, which will concern us for the bulk of the section. Fix a smooth, contractible manifold $M$, which we assume to be four dimensional.  By a $U(1)$ \emph{gauge field} on $M$, we mean a smooth one-form $A_a$; following the notation of \citep[\S 3.1]{Nguyen+etal}, a \emph{gauge transformation} is a map from gauge fields to gauge fields of the form
\[
A_a\mapsto A_a + g^{-1}d_a g
\]
where $g:M\rightarrow U(1)$ is a smooth map.\footnote{To unpack this equation: by $d_a g$, we mean the pushforward map along $g$ defined at each point, which, at each $p\in M$, is a map from $T_pM$ to $T_{g(p)}U(1)$.  Then $g^{-1}$ is the pushforward along the translation on $U(1)$ determined by the inverse of the group element $g(p)$, yielding an element of the tangent space at the identity of $U(1)$, i.e., an element of the Lie algebra of $U(1)$ (which happens to be $\mathbb{R}$).  Thus, $g^{-1}d_a g$ is a (closed) one-form on $M$.}  Observe that on this definition, since gauge transformations are parameterized by maps $g$, all gauge fields are related to themselves by gauge transformations that are constant maps from $M$ to $U(1)$---that is, there are gauge transformations that are non-trivial `automorphisms' of gauge fields.

Nguyen et al. are concerned with the relationship between several different categories that one might define to characterize the structure of such gauge fields.\footnote{In all of these categories, following Nguyen et al., we `fix' $M$.  For some purposes, one might wish to include diffeomorphisms acting on $M$ among the morphisms of the categories, but nothing is lost for present purposes by neglecting them.}  
\begin{itemize}
\item $\mathcal{C}_A$: objects are gauge fields $A_a$; morphisms are gauge transformations;
\item $\mathcal{S}_{[A]}$: objects are equivalence classes $[A]$ of gauge fields under gauge transformations;  morphisms are identity maps;
\item $\mathcal{S}_{A}$: objects are gauge fields; morphisms are identity maps;
\item $\mathcal{E}_A$: objects are gauge fields; morphisms are equivalence classes of gauge transformations, where $g\sim h $ if $g^{-1}d_a g - h^{-1}d_a h=\mathbf{0}$.
\end{itemize}
The category $\mathcal{S}_A$ is what one gets if one takes each gauge field on $M$ to represent a distinct possible situation; any non-identical, gauge-related gauge fields are inequivalent by the lights of this category.  The category $\mathcal{C}_A$ is what one gets if one takes gauge fields to represent possible situations, but where every gauge transformation represents an `equivalence' of gauge fields.  

The categories $S_{[A]}$ and $\mathcal{E}_A$ are categorically equivalent. Both express the idea that any two gauge fields related by gauge transformations are equivalent to one another, such that it is only the equivalence classes of gauge fields that have physical significance.  They are also both equivalent to yet another category, $\mathcal{S}_F$, whose objects are smooth two-forms $F_{ab}$ on $M$ satisfying $d_aF_{bc}=\mathbf{0}$.  In the context of electromagnetism, such tensors represent the electromagnetic field; they are related to gauge fields by the equation $F_{ab}=d_a A_b$, with any two gauge fields related by a gauge transformation giving rise to the same electromagnetic field.  So one can think of $S_{[A]}$ and $\mathcal{E}_A$ as representing the theory that says it is the electromagnetic fields on $M$ that represent distinct possible situations, with different gauge fields representing different situations only if they give rise to different electromagnetic fields.  Thus, $\mathcal{S}_{[A]}$ and $\mathcal{E}_A$ correspond to a widespread view that it is the electromagnetic fields, and not the gauge fields directly, that have physical significance in electromagnetism.

The category $\mathcal{S}_A$, meanwhile, has more structure that either of these, in the sense defined above: there is a functor from $\mathcal{S}_A$ to $\mathcal{S}_{[A]}$ that preserves empirical content, and which is not full.\footnote{This is because there are (gauge-equivalent) objects $A_a$ and $A'_a$ of $\mathcal{S}_A$ that are mapped to the same object $[A]$ of $\mathcal{S}_{[A]}$, but which have no arrows between them that could map to the identity on $[A]$.}  (It is faithful and essentially surjective.)  This functor takes gauge fields $A_a$ to their equivalence classes under gauge transformations, and takes all arrows to identities.  It is this relationship that is emphasized in \citep{WeatherallUG}, to capture the idea that a theory in which one takes different gauge-related gauge fields to be inequivalent has, in a precise sense, more structure than a theory in which one takes gauge-related gauge fields to be equivalent---or, in light of the equivalence to $\mathcal{S}_F$ already noted, that a theory that distinguishes gauge fields has more structure than one that distinguishes (only) electromagnetic fields.  It was in this sense that Weatherall claimed to give a precise characterization of the claim that electromagnetism formulated using gauge fields has `surplus structure': it is because there is another formulation available with less structure, but with the same empirical consequences.\footnote{We remark that there is also a functor going in the opposite direction that (using Choice) chooses, from each equivalence class $[A]$, a representative $A_a$.  It is interesting to note that this functor is full and faithful, because every arrow is mapped to an identity arrow, and no two objects have more than one arrow between them; but not essentially surjective, because each equivalence class is identified with a single representative.  So this functor forgets forgets property---not structure \emph{or} stuff.  To see what is going on here, note that what this functor is doing is associating with each equivalence class a single, preferred representative.  But from the point of view of $\mathcal{S}_A$, there are many other fields around that do not get mapped to, representing physical possibilities that are inequivalent to those in the image of the functor, but which do not correspond to any possibility represented in $\mathcal{S}_{[A]}$, according to that functor.  We can think of the property that is forgotten as the property of being the (privileged) representative of an equivalence class (or the `one true gauge').}

But what about $\mathcal{C}_A$?  As Nguyen et al. show, $\mathcal{C}_A$ has surplus* structure relative $\mathcal{E}_{A}$---that is, there is a functor $\tau: \mathcal{C}_A\rightarrow \mathcal{E}_{A}$ that forgets stuff and preserves empirical content.\footnote{It was essentially this functor that \citet{WeatherallUG} mistakenly claimed forgot nothing; see \citet{WeatherallErratum}.}  It is this functor, they argue, that supports their claim that there is a sense in which Yang-Mills theory has something `surplus', even after one takes gauge transformations to be equivalences.  They write, `This ... precisification of `surplus' structure allows us to define \emph{surplus* structure} as the \emph{stuff} that is forgotten by the functor $\tau$,', and then go on to say: `A (gauge) theory contains the \emph{stuff} that is forgotten by $\tau$ (namely the non-trivial automorphisms of the gauge fields and the result of concatenating them with the morphisms already contained in $\mathcal{E}_A$)' (p. 11).  They then proceed to argue that it is really this relationship between $\mathcal{C}_A$ and $\mathcal{E}_A$, and not that between $\mathcal{E}_A$ and $\mathcal{S}_A$, that captures the physically salient sense in which a gauge theory has surplus structure.  As they write, 
\begin{quote}\singlespacing [W]e are interested in a notion of `surplus' that is possessed by theories which take gauge fields to be representationally equivalent (and which represent this by means of gauge transformations between gauge fields); thus $\mathcal{C}_A$ is our candidate for such a theory and `surplus' is characterised by the \emph{stuff}-forgetting functor $\tau:\mathcal{C}_A\rightarrow S_{[A]}$....  By contrast, Weatherall's notion of `surplus' applies to a theory that does not represent gauge fields as representationally equivalent, namely $\mathcal{S}_A$....\end{quote}
Thus, even \emph{after} one has taken all of the gauge fields related by gauge transformations to be equivalent, one \emph{still} as a theory with some surplus---namely, the surplus gauge transformations!

There are a few remarks to make, here.  First, as one might expect given our discussion in section \ref{sec:W}, $\mathcal{C}_A$ and $\mathcal{E}_A$ are also related in another salient way: there is also a functor $K:\mathcal{E}_A\rightarrow \mathcal{C}_A$ that forgets structure (and preserves empirical content).  Thus, the difference between Weatherall's account and Nguyen et al.'s account is not just that they are comparing $\mathcal{E}_A$ to different categories and getting different accounts.  In fact, these two criteria for when one theory has more structure than another yield precisely \emph{opposite} verdicts: one says that $\mathcal{C}_A$ has more structure; the other says $\mathcal{E}_A$ does.  If one adopts the view we have defended here, then, moving to $\mathcal{C}_A$ involves forgetting further structure, even relative to $\mathcal{E}_A$.  

So it would seem we have two different criteria, giving opposite verdicts.  Which, if either, is right?  As things stand, it is difficult to see what is at stake in the disagreement.  The reason is that, as things have been set up so far, both $\mathcal{C}_A$ and $\mathcal{E}_A$ take precisely the same gauge fields to be equivalent: namely, the ones related by gauge transformations.  So if one were pressed to say what structure each of these categories attributes to the world, it would be tempting to say `equivalence classes of gauge fields under gauge transformation.'  And yet, on both  criteria of structural comparison under consideration, these theories are not equivalent.  The difference between the categories---a difference that \emph{both} criteria are tracking---concerns the additional morphisms, such as the non-trivial automorphisms, of $\mathcal{C}_A$.  But from the perspective of the structure we attribute to the world in the models of these theories, it is difficult to see what these additional transformations reveal.  After all, they are (all) maps of the form $A_a\mapsto A_a + \mathbf{0}$.\footnote{One might even worry about the following proposal: suppose we have a theory, and we would like another theory with `less structure'.  We could simply stipulate that every model of the theory is equivalent to itself in more ways, by introducing trivial maps.  For instance, in a model of general relativity, consider the new metric automorphisms which are maps of the form $g_{ab}\mapsto g_{ab} + n\mathbf{0}$, for all $n$.  Suddenly metrics have a new automorphism group!}

To see what is going on, here, we need to think of these categories---or really, the theory of electromagnetism---from a different perspective.\footnote{For a detailed overview of this perspecitve, written for philosophers, see \citep{WeatherallFBYMGR}; see also \citep{Bleecker} and \citep{Palais} for excellent mathematical treatments of the subject.}  On this alternate approach, a gauge field is not conceived as a one-form on $M$; instead, it is a principal connection $\omega_{\alpha}$ on a $U(1)$ principal bundle $P\xrightarrow{\pi} M$ over $M$.\footnote{A principal $G$ bundle, for some Lie group $G$, is a smooth surjective map $P\xrightarrow{\pi} M$, where $M$ and $P$ are smooth manifolds with the following property: there is a smooth, free, fiber-preserving right action of $G$ on $P$ such that given any point $p\in M$, there exists a neighborhood $U$ of $p$ and a diffeomorphism $\zeta:U\times G\rightarrow \wp^{-1}[U]$ such that for any $q\in U$ and any $g,g'\in G$, $\zeta(q,g)g'=\zeta(q,gg')$.  A (global) \emph{section} of a principal bundle is a smooth map $\sigma:M\rightarrow P$ satisfying $\pi\circ\sigma=1_M$.  A \emph{principal connection} on $\pi$ is a smooth Lie-algebra-valued one form $\omega^{\mathfrak{A}}{}_{\alpha}$ on $P$ satisfying certain further conditions, including that $\omega^{\mathfrak{A}}{}_{\alpha}$ be surjective on the Lie algebra.  (Here the lowered Greek index indicates action on tangent vectors to $P$ and the raised capital fraktur index indicates membership in the Lie algebra of $G$, $\mathfrak{g}$.  Since the Lie algebra of $U(1)$ is $\mathbb{R}$, we drop the fraktur index when discussing principal connections on $U(1)$ bundles.)}  A gauge field in the earlier sense, $A_a$, arises as a representation of $\omega_{\alpha}$ on $M$, relative to a choice of (global) section, $\sigma:M\rightarrow P$, by $A_a=\sigma^*(\omega_{\alpha})$.  Gauge transformations, meanwhile, can be identified with changes of sections. Since $M$ is contractible, there is a unique principal $U(1)$ bundle over $M$, and so all of the gauge fields under consideration are principal connections on that unique principal bundle, represented relative to different sections. 

From this perspective, the category $\mathcal{C}_A$ is (isomorphic to) the category whose objects are principal connections $\omega_{\alpha}$ on $P$ and whose morphisms are vertical principal bundle (auto)morphisms that preserve $\omega_{\alpha}$.\footnote{A \emph{vertical principal bundle automorphism} on $\pi$ is a diffeomorphism $\Psi:P\rightarrow P$ such that (a) $\pi\circ\Psi = \pi$ and (b) for any $x\in P$ and $g\in G$, $\Psi(xg)=\Psi(x)g$.  We remark that although these maps are automorphisms on the principal bundle, they are not necessarily automorphisms once one fixes a connection.}  The `extra' morphisms in $\mathcal{C}_A$ can then be seen as maps that do not act trivially after all: they are symmetries of the connection $\omega_{\alpha}$ under (non-trivial) transformations of the bundle $\pi$.  The category $\mathcal{E}_A$, on the other hand, is (isomorphic to) what results if one adds, to this principal bundle, some further `rigidifying' structure, which breaks these symmetries. There are several equivalent ways to characterize the sort of structure that would do this, but one natural candidate is a fixed, global trivialization of the bundle, i.e., a fixed diffeomorphism $\zeta:P\rightarrow M\times G$ such that for any $p\in M$ and any $g,g'\in G$, $\zeta(q,g)g'=\zeta(q,gg')$.  (The fact that such a global trivialization exists is a consequence of the contractibility of $M$.)  Another way of thinking about this structure is as fixing a choice of identity in the `fiber' $\pi^{-1}[p]$ over each point of $M$, thus fixing the way in which the fiber realizes the Lie group structure.

It is perhaps worth emphasizing two related points about principal bundles at this stage, to clarify the remarks just made.  First, although principal bundles, like all fiber bundles, are required to be `locally trivial' in the sense that they admit local trivializations, these have a status similar to coordinate charts on manifolds, and there are no `privileged' trivializations.  Fixing a global trivialization is analogous to choosing a global coordinate system on a manifold.  The second point is that in general, the fibers of a principal bundle are `$G$ torsors', which are manifolds diffeomorphic to $G$ on which $G$ acts, but which are not themselves Lie groups because they do not have a group structure (and do not act on themselves).  So fixing an origin for each fiber can be thought of as endowing the $U(1)$-torsors of $\pi$ with Lie group structure.  Both of these remarks are (complementary and compatible) ways of pinning down what, exactly, changes when one moves from $\mathcal{C}_A$ to $\mathcal{E}_A$: the latter is the category of principal connections on principal bundles endowed with something that, intuitively, seems like further structure: a global trivialization, akin to a global coordinate system; or Lie group structure on each fiber.  On the other hand, as the discussion in section \ref{sec:W} would lead one to expect, $\mathcal{C}_A$, which has less structure, has more stuff---reflecting the representational redundancy (freedom) afforded by the many ways of endowing the bundle with this structure.


The upshot of this discussion is that we can see the lessons of the previous sections in practice. We have a sense in which the theory characterized by $\mathcal{C}_A$ has less structure than that of $\mathcal{E}_A$, and correspondingly more representational redundancy, in the first sense discussed above---namely, that captured by surplus* structure (or, surplus stuff). Conversely, we can think of $\mathcal{E}_A$ as also having representational redundancy in the third sense, i.e., that of surplus structure, because by having a fixed global trivialization, this theory treats different choices of trivialization as being non-isomorphic, which are nonetheless observationally equivalent.  Finally, we remark that \emph{both} theories have representational redundancy in the second sense, because in both cases there are isomorphic models that may all represent a given situation equally well; but this sense of representational redundancy is irrelevant to which theory has more structure.

There might be a lingering dissatisfaction at this stage.  Above, following Nguyen et al., we described the categories $\mathcal{E}_A$ and $S_{[A]}$ without ever mentioning any principal bundles---much less the extra `structure' of a global trivialization.  So in what sense should we think of these categories as representing theories that invoke such structure?  The answer is subtle.  When we defined gauge fields above, we introduced them as one-forms on $M$.  One-forms form a vector space at each point, with the zero covector as the origin.  Once we recognize that these gauge fields are really principal connections, however, we can see a sense in which they should form an affine space at each point, rather than a vector space.  One can see this by observing, for instance, that there is no `zero connection', because every principal connection must be a surjective linear transformation; and the addition of two principal connections is not necessarily a principal connection.  There do, however, exist connections $\omega_{\alpha}$ and sections $\sigma:M\rightarrow P$ such that $\sigma^*(\omega_{\alpha})=\mathbf{0}$; and relative to such a choice of connection, the space of connections at each point takes on the structure of a vector space (because $\omega_{\alpha}$ fixes an origin), elements of which can be put into one-to-one correspondence with one-forms on $M$, via $\sigma$. Thus we can see a certain sense in which the way we set up the theory in the first place, associating gauge fields with one-forms on $M$, already relied on a choice of background structure; from this perspective, the `extra' arrows of $\mathcal{C}_A$ are a way of washing out this vector space structure on connections.

We now turn to the second argument that Nguyen et al. give, which we treat much more briefly because we do not object to its substance.  The second argument is that in fact, the surplus* structure that $\mathcal{C}_A$ has over $\mathcal{S}_{[A]}$ is an essential feature of electromagnetism. The reason---restated in the terms of the discussion above---is that a formulation of electromagnetism that can treat the full range of possible gauge field configurations on topologically non-trivial (i.e., non-contractible) spacetime manifolds $M$ in a suitably local way must be able to represent the principal connections that one can define on principal bundles for which no global trivialization exists---that is, non-trivial principal bundles.  If no global trivialization exists, it is not possible to fix a global trivialization and identify the space of principal connections  on the bundle with the one-forms on the base space. This problem does not arise for contractible manifolds, because the only principal bundles over such manifolds do admit global trivializations.   This makes precise, in a somewhat different language from that used by Nguyen et al, the sense in which equivalence classes of one-forms on $M$ under gauge transformation cannot adequately represent the full richness of Yang-Mills theory. 

Thus, we agree that the surplus* structure captured by $\mathcal{C}_A$---that is, the representational redundancy afforded by a principal bundle formulations of Yang-Mills theory---is necessary to capture the full richness of the theory.  What we disagree about is the interpretation of this conclusion.  The moral that \cite{Nguyen+etal} draw is that the morphisms of a category, as well as the objects (the models), can feature in the representational content of a theory; in the case of $\mathcal{C}_A$, they represent the ways in which local fields (given by the models of $\mathcal{C}_A$) can be composed to give global systems. On this view, the reason $\mathcal{C}_A$ is superior to $\mathcal{E}_A$ is that the \emph{prima facie} `surplus', namely the morphisms of $\mathcal{C}_A$ that contribute to the representational redundancy of the theory, have representational significance.  Hence their title claim: surplus structure is not superfluous.

To the contrary, on the view we have defended here, the reason $\mathcal{C}_A$ is adequate, but $\mathcal{E}_A$ is not, is precisely that the former has \emph{less} structure than the latter, which affords it greater representational capacities; put another way, the structure invoked to get from $\mathcal{C}_A$ to $\mathcal{E}_A$ cannot be consistently imposed in all cases of physical interest.  From this perspective, the problem with $\mathcal{E}_A$ is analogous to the problem faced by someone who insists that there should be some preferred global coordinate systems in special relativity.  One can define such a thing in that context; but when one moves to general relativity, one cannot recover the full richness of the theory if one insists on always having global coordinate systems, because some models of general relativity do not admit such systems.  The upshot is that surplus \emph{stuff} may well not be superfluous, but surplus \emph{structure} is not only superfluous, but in some cases it is a barrier to capturing the full richness of a theory.

This conclusion is in many ways irenic: in the end, we agree with Nguyen et al. on the principal conclusion of their paper, that one should take $\mathcal{C}_A$ to be the best categorical representation of the structure of electromagnetism, even in the contractible case.  We disagree only on our route to the conclusion, namely whether it goes via a comparison of `stuff' or of `structure'.  But the difference in perspective matters to the rhetorical posture in \citep{Nguyen+etal}.  Their argument is motivated by an alleged puzzle, which is: `how can `surplus' structure be an essential feature of a theory?' (p. 12).\footnote{See also p. 2: `How can `redundancy' be an essential feature of a theory?'  This formulation is more congenial to our perspective here.}  But if $\mathcal{C}_A$ has \emph{less} structure than $\mathcal{E}_A$, then this puzzle never arises, for it is easy to see how eliminating structure can be essential for a theory, particularly when that structure can only be defined for a small subset of the possible models of the theory. 

\section{The Hole Argument}\label{sec:holes}

We now return to the hole argument in light of the distinctions regarding surplus structure and representational redundancy drawn above. As we noted in the Introduction, one of the issues that the hole argument is sometimes taken to highlight is that the standard formulation of general relativity, using tensor fields on a smooth manifold, has `too much structure'.  The hole argument, recall, uses the fact that given a model of general relativity, $(M, g_{ab})$,\footnote{Here $M$ is a smooth, four-dimensional manifold, which we assume to be Hausdorff and paracompact; and $g_{ab}$ is a smooth, Lorentz-signature metric $g_{ab}$ defined on $M$.  For further details on the mathematical background of general relativity, see \citet{Wald} or \citet{MalamentGR}. Our discussion in what follows depends only on the sorts of mathematical facts that are usually at issue in the literature on the hole argument.} one can construct another model $(M,g_{ab}')$ through a diffeomorphism $\psi: M \rightarrow M$ on the manifold, where $g_{ab}'$ is defined by the pushforward map determined by the diffeomorphism, $g_{ab}' = \psi_*(g_{ab}) $. If the diffeomorphism does not act as the identity everywhere, then these models agree on all observable structure and yet disagree at certain points on the value of the metric. And so---the argument goes---there must be some `surplus structure' in the standard formulation that is physically insignificant. 

However, as we have now seen, there are different ways one might understand `surplus structure'. One sense is that a (formulation of a) theory has surplus structure if there is another formulation of the theory and a functor from the first theory to the second, represented as categories, that is not full (and preserves empirical content).  As \citet{WeatherallUG} argues, the hole argument does not reveal that general relativity has surplus structure in this sense.  For it to do so, it would need to be the case that the hole argument generated models of general relativity that were empirically indistinguishable but \emph{not isomorphic} by the lights of the ambient mathematical theory (i.e., the theory of Lorentzian manifolds).  If this were to occur, one might hope to move to another formulation of the theory on which these models \emph{were} isomorphic.  But this is precisely what the hole argument does not do.\footnote{Of course, some authors have taken the hole argument to do something like what is described here.  But as \citet{WeatherallHoleArg} argues, this is chimerical: to get to the conclusion that the hole argument generates empirically equivalent but non-isomorphic possibilities, one uses the identity map on the manifold to compare particular points on the manifold. Under such a comparison, the models are not equivalent, either representationally (because it does not give rise to an isomorphism) or observationally.} Therefore, the standard formulation does not have surplus structure in this sense, and, \emph{a fortiori}, the hole argument does not reveal otherwise. 

But what about the other sense of `surplus structure'---viz., surplus* structure or representational redundancy in the first sense of section \ref{sec:Ladyman}, as described by \citet{Nguyen+etal}?  First, there is a sense in which general relativity, as ordinarily formulated, has surplus* structure.  Of course, just as with surplus structure, surplus* structure is a relative notion: one theory has surplus* structure relative to another if there is a functor from the first to the second that forgets stuff (and preserves empirical content).  But we can develop a heuristic for identifying when this is likely to happen, similar to the one we gave above for surplus structure: we will say that a theory, represented as a category, has surplus* structure if there exist models $A$ and $B$ of the theory and isomorphisms $f,g:A\rightarrow B$ such that $f\neq g$.  In particular, it suffices for a theory to have surplus* structure in this sense if any of its models has a non-trivial automorphism group.  This heuristic captures the idea that the theory has `stuff' to forget, in the form of `extra' arrows; it also captures the idea of the theory having representational redundancy, since as we saw, that is signaled by symmetries of the models of the theory.  And we can also see immediately that general relativity has surplus* structure by this criterion: simply observe that Minkowski spacetime---a model of general relativity---has non-trivial symmetries (such as translations or Lorentz boosts).

This argument shows a sense in which general relativity does have surplus* structure---but it does not involve the hole argument.  In fact, the maps involved in the hole argument---that is, isometries generated by diffeomorphisms from a manifold to itself---in general are \emph{not} automorphisms of any spacetime, in the sense just described, because in general $\psi_*(g_{ab})\neq g_{ab}$.  Nor is it the case that, in general, isometric pairs of spacetimes $(M,g_{ab})$ and $(M,\psi_*(g_{ab}))$ generated by the hole argument are related by any further isometries.  So not only does the hole argument play no role in the argument just given that general relativity has surplus* structure, it also does not generate the sort of mappings that, we just argued, should be taken to signal surplus* structure.

Nonetheless, there are connections between the hole argument and surplus* structure.  To see this connection, observe that the hole argument \emph{does} reveal that general relativity has some representational redundancy, in the sense that, given a physical situation that can be modeled by general relativity at all, the hole argument shows that there always exist nondenumerably many isometric spacetimes that one can choose between to model that situation.  This is not representational redundancy in the sense captured by surplus* structure, but rather representational redundancy in the \emph{second} sense of section \ref{sec:Ladyman}.  

As we suggested there, this sort of representational redundancy may be seen to arise from surplus* structure at a different level.  In particular, what the hole argument exploits is the fact that (bare) \emph{manifolds} have surplus* structure, as representations of spacetime.  Consider the category \textbf{4Man} whose objects are smooth four dimensional manifolds and whose arrows are diffeomorphisms.  This category has a rich structure of automorphism groups, signaling, by the heuristic above, that has surplus* structure.  And indeed, given, say, the manifold $\mathbb{R}^4$, there are many ways in which one could use that manifold to represent the events of, say, our own universe (assuming that our universe is topologically simple).  Indeed, here we find ourselves in a situation strikingly similar to that of the person trying to use a two dimensional vector space (with no further structure) to represent the cardinal directions.  Any point at all of $\mathbb{R}^4$ can be used, equally well, to represent `here-now'; likewise, any distinct point can be use to represent `over there ten minutes ago'; and so on.  

One way of understanding what the hole argument is doing, then, is taking different ways of exercising this freedom to represent events in space and time with a given manifold $M$, and showing that different choices of how to represent (say) `here-now' give rise to distinct, though isometric, spacetimes once we include metrical structure.\footnote{The argument in \citep{WeatherallHoleArg} concerns just this issue---or rather, a possible misunderstanding concerning how to understand this `freedom'.}  This is just as in the vector space case above, where different choices of vectors to represent `North' and `East' give rise to distinct but isomorphic representations of the cardinal directions.

One should be reluctant to infer too much from these reflections, however.  The reason is that---as we remarked in section \ref{sec:Ladyman}---this sort of representational redundancy will \emph{always} be present for any theory formulated with a set theoretic semantics.\footnote{\citet{Rynasiewicz} draws a very similar moral regarding the hole argument, relating it to Putnam's famous permutation argument.} The fact that bare manifolds have surplus* structure that Lorentzian manifolds lack does not reflect anything deep about the theory; instead, it merely reflects the fact that we tend to build mathematical objects out of other mathematical objects with less structure.  

What these reflections \emph{do} serve to illustrate is rather how surplus* structure relates to ideas that have a long provenance in the philosophy of physics literature.  The hole argument does make use of surplus* structure, and many philosophers have taken the existence of this surplus* structure, or at least consequences of its existence, to suggest strong morals regarding the adequacy of the standard formalism of general relativity.  But as we hope we have shown here, the fact that manifolds have surplus* structure, of the sort exploited in the hole argument, should not be taken to imply that manifolds, or spacetimes, or general relativity more generally, have surplus structure.  To the contrary, that manifolds have surplus* structure suggests that they have \emph{less} structure than, say, Lorentzian manifolds.\footnote{Or, we hasten to add, Einstein algebras, since the latter are, in a precise sense, equivalent to relativistic spacetimes \citep{Rosenstock+etal}.}

There is a certain irony to all of this.  Once we are clear about the difference between surplus structure and surplus* structure, and we see that the hole argument exploits the surplus* structure of manifolds, rather than any surplus structure of manifolds or of relativistic spacetimes, we can now ask what it would mean to adopt a methodological dictum to `minimize' either surplus structure or surplus* structure.  Since it is not clear that general relativity \emph{has} surplus structure---and indeed, as we have argued, there are good reasons to think it does not---the dictum that says `minimize surplus structure' would recommend keeping the standard formalism.  As we have seen, however, general relativity \emph{does} have surplus* structure.  And this surplus* structure can be eliminated, basically by adding further structure to relativistic spacetimes to `rigidify' them, in the sense of removing any non-trivial automorphisms.  One way of doing this would be to fix a global labeling system, so that each point is given a unique label.\footnote{Note the echoes of the move from $\mathcal{C}_A$ to $\mathcal{E}_A$ as described in the previous section.  Note that one cannot generally introduce a global \emph{coordinate system}, in the sense of a smooth map from a generic four dimensional manifold to $\mathbb{R}^4$, but one can always assign unique labels to points, for instance by fixing some (non-smooth) map to $\mathbb{R}^4$} 

And this brings us to the irony, which is that this sort of model of spacetime, where one has a Lorentzian manifold endowed with an `individuating field' has been proposed, for instance by \citet{StachelSS} and \citet{WeatherallHoleArg,WeatherallStein}, as a way of capturing the structure that the `manifold substantivalist' wishes to express, namely, that locations, or points, of spacetime have some ontological status independent of or prior to the events that occur there.  And so, minimizing surplus* structure, far from \emph{eliminating} the structure that the substantivalist wished to endorse, leads us to \emph{add} that structure to spacetime.  And there are good reasons not to do this---not least of which is that the hole argument shows that a theory that posited \emph{this} structure would not be deterministic, or at least, the evolution of the world would not be uniquely determined by Einstein's equation up to isomorphism.  And so we find another connection between surplus* structure and the hole argument: in the context of general relativity, the hole argument provides strong reasons \emph{not} to minimize surplus* structure. 

\section{Conclusion}

We have studied two proposals for how to use category theory to make precise the idea of a physical theory having `surplus structure': those of \citet{WeatherallUG} and \citet{Nguyen+etal}.  We have argued, by looking at simple examples, that surplus* structure in the sense of Nguyen et al. does not generally correspond to `surplus structure' in at least one intuitive sense common in mathematics and philosophy of physics.  To the contrary, we argued, having surplus* structure is generally associated with a kind of representational redundancy that signals having \emph{less} structure---and to remove representational redundancy in this sense requires adding structure to a theory. Thus, although we agree with Nguyen et al. that Yang-Mills theory has surplus* structure, and more, that it is essential for Yang-Mills theory to have surplus* structure to accommodate the full range of physical situations to which one would like to apply it, we do not think this conclusion is in tension with the idea that one should wish to minimize structure in physical theories.  To the contrary, we argue, the reason Yang-Mills theory is able to do the work that Nguyen et al. highlight is that formulations of the theory with less surplus* structure require one to fix structure globally in a way that cannot be done consistently in all cases.  Hence it is the fact that removing surplus* structure amounts to \emph{adding} structure, in the sense of \citep{WeatherallUG}, that vitiates these alternative formulations of Yang-Mills theory that Nguyen et al. (correctly) argue are inadequate.

We then applied these morals to the hole argument, arguing that the differences between surplus structure and surplus* structure helps clarify both why the hole argument does not reveal that the standard formalism of general relativity is inadequate, and also why one might have thought it did.  The reason for the latter is that the hole argument does invoke the surplus* structure, or representational redundancy, of manifolds as representations of events in space and time.  But this should not be taken to signal a problem with the standard formalism---nor should it motivate moving to a different formulation with less surplus* structure than manifolds, because to do so would amount to fixing \emph{extra} structure, along the lines of what the manifold substantivalist endorses.  This helps clarify the competing intuitions in the literature on the hole argument, but also leads to the ironic situation that the sort of representational redundancy that some philosophers have taken to signal a problem with general relativity is precisely what one gets when one adopts a formalism that avoids surplus* structure.

\section*{Acknowledgments}
This material is partially based upon work produced for the project “New Directions in Philosophy of Cosmology”, funded by the John Templeton Foundation under grant number 61048.  We are grateful to Thomas Barrett, James Ladyman, and Nic Teh for helpful discussions and suggestions as we prepared this paper.  

\bibliographystyle{elsarticle-harv}
\bibliography{holes}

\begin{thebibliography}{43}
\expandafter\ifx\csname natexlab\endcsname\relax\def\natexlab#1{#1}\fi
\expandafter\ifx\csname url\endcsname\relax
  \def\url#1{\texttt{#1}}\fi
\expandafter\ifx\csname urlprefix\endcsname\relax\def\urlprefix{URL }\fi

\bibitem[{Baez et~al.(2004)Baez, Bartel, and Dolan}]{Baezetal}
Baez, J., Bartel, T., Dolan, J., 2004. Property, structure, and stuff.
  Available at:
  \url{http://math.ucr.edu/home/baez/qgspring2004/discussion.html}.

\bibitem[{Baez and Shulman(2010)}]{Baez+Shulman}
Baez, J., Shulman, M., 2010. Lectures on n-categories and cohomology. In: Baez,
  J.~C., May, J.~P. (Eds.), Towards Higher Categories. Springer, Dordrecht, pp.
  1--68.

\bibitem[{Bain(2003)}]{Bain}
Bain, J., 2003. Einstein algebras and the hole argument. Philosophy of Science
  70~(5), 1073--1085.

\bibitem[{Barrett(2014)}]{BarrettCM1}
Barrett, T., 2014. On the structure of classical mechanics. The British Journal
  for the Philosophy of Science 66~(4), 801--828.

\bibitem[{Barrett(2015)}]{BarrettSTS}
Barrett, T., 2015. Spacetime structure. Studies in History and Philosophy of
  Modern Physics 51, 37--43.

\bibitem[{Barrett(2018)}]{BarrettSS}
Barrett, T.~W., 2018. What do symmetries tell us about structure? Philosophy of
  Science 85~(4).

\bibitem[{Bleecker(1981)}]{Bleecker}
Bleecker, D., 1981. Gauge Theory and Variational Principles. Addison-Wesley,
  Reading, MA, reprinted by Dover Publications in 2005.

\bibitem[{Earman(1986{\natexlab{a}})}]{EarmanPD}
Earman, J., 1986{\natexlab{a}}. A Primer on Determinism. D. Reidel, Dordrecht.

\bibitem[{Earman(1986{\natexlab{b}})}]{Earman1986}
Earman, J., 1986{\natexlab{b}}. Why space is not a substance (at least not to
  first degree). Pacific Philosophical Quarterly 67~(4), 225--244.

\bibitem[{Earman(1989{\natexlab{a}})}]{Earman1989}
Earman, J., 1989{\natexlab{a}}. Leibniz and the absolute vs. relational
  dispute. In: Rescher, N. (Ed.), Leibnizian Inquiries. A Group of Essays.
  University Press of America, Lanham, MD, pp. 9--22.

\bibitem[{Earman(1989{\natexlab{b}})}]{EarmanWEST}
Earman, J., 1989{\natexlab{b}}. World Enough and Space-Time. The MIT Press,
  Cambridge, MA.

\bibitem[{Earman and Norton(1987)}]{Earman+Norton}
Earman, J., Norton, J., 1987. What price spacetime substantivalism? the hole
  story. The British Journal for the Philosophy of Science 38~(4), 515--525.

\bibitem[{Field(1984)}]{Field}
Field, H., 1984. Can we dispense with space-time? In: PSA: Proceedings of the
  Biennial Meeting of the Philosophy of Science Association. Vol. 1984.
  Philosophy of Science Association, pp. 33--90.

\bibitem[{Fletcher(2020)}]{Fletcher}
Fletcher, S.~C., 2020. On representational capacities, with an application to
  general relativity. Foundations of PhysicsThis volume. DOI:
  10.1007/s10701-018-0208-6.

\bibitem[{Friedman(1983)}]{Friedman}
Friedman, M., 1983. Foundations of Space-Time Theories: Relativistic Physics
  and Philosophy of Science. Princeton University Press, Princeton, NJ.

\bibitem[{Geroch(1972)}]{GerochEA}
Geroch, R., 1972. Einstein algebras. Communications in Mathematical Physics 26,
  271--275.

\bibitem[{Gilton(2018)}]{Gilton}
Gilton, M. J.~R., 2018. Could charge and mass be universal properties?,
  unpublished ms.

\bibitem[{Halvorson(2012)}]{Halvorson}
Halvorson, H., 2012. What scientific theories could not be. Philosophy of
  Science 79~(2), 183--206.

\bibitem[{Healey(2007)}]{Healey}
Healey, R., 2007. Gauging What's Real: The Conceptual Foundations of
  Contemporary Gauge Theories. Oxford University Press, New York.

\bibitem[{Leinster(2014)}]{Leinster}
Leinster, T., 2014. Basic Category Theory. Cambridge University Press,
  Cambridge.

\bibitem[{{Mac Lane}(1998)}]{MacLane}
{Mac Lane}, S., 1998. Categories for the Working Mathematician, 2nd Edition.
  Springer-Verlag, New York.

\bibitem[{Malament(2012)}]{MalamentGR}
Malament, D.~B., 2012. Topics in the Foundations of General Relativity and
  {Newton}ian Gravitation Theory. University of Chicago Press, Chicago.

\bibitem[{Nguyen et~al.(2018)Nguyen, Teh, and Wells}]{Nguyen+etal}
Nguyen, J., Teh, N.~J., Wells, L., 2018. Why surplus structure is not
  superfluous. British Journal for Philosophy of ScienceForthcoming.

\bibitem[{Norton(1984)}]{Norton1984}
Norton, J.~D., 1984. How einstein found his field equations: 1912–1915.
  Historical Studies in the Physical Sciences 14, 253--316.

\bibitem[{Palais(1981)}]{Palais}
Palais, R.~S., 1981. The Geometrization of Physics. Institute of Mathematics,
  National Tsing Hua University, Hsinchu, Taiwan, available at
  http://vmm.math.uci.edu/.

\bibitem[{Rosenstock(2018)}]{Rosenstock}
Rosenstock, S., 2018. Functoriality and the structure of data, unpublished
  manuscript.

\bibitem[{Rosenstock et~al.(2015)Rosenstock, Barrett, and
  Weatherall}]{Rosenstock+etal}
Rosenstock, S., Barrett, T., Weatherall, J.~O., 2015. On {Einstein} algebras
  and relativistic spacetimes. Studies in History and Philosophy of Modern
  Physics 52B, 309--316.

\bibitem[{Rovelli(2006)}]{RovelliDisappearance}
Rovelli, C., 2006. The disappearance of space and time. In: Dieks, D. (Ed.),
  The Ontology of Spacetime. Elsevier, Amsterdam, pp. 25--36.

\bibitem[{Rynasiewicz(1992)}]{Rynasiewicz1992}
Rynasiewicz, R., 1992. Rings, holes and substantivalism: On the program of
  {L}eibniz algebras. Philosophy of Science 59~(4), 572--589.

\bibitem[{Rynasiewicz(1994)}]{Rynasiewicz}
Rynasiewicz, R., 1994. The lessons of the hole argument. The British Journal
  for the Philosophy of Science 45~(2), 407--436.

\bibitem[{Smolin(2000)}]{SmolinThreeRoads}
Smolin, L., 2000. Three roads to quantum gravity. Basic books, New York.

\bibitem[{Stachel(1989)}]{Stachel}
Stachel, J., 1989. Einstein's search for general covariance, 1912--1915. In:
  Howard, D., Stachel, J. (Eds.), Einstein and the History of General
  Relativity. Birkhauser, Boston, pp. 62--100.

\bibitem[{Stachel(1993)}]{StachelSS}
Stachel, J., 1993. The meaning of general covariance. In: Earman, J., Janis,
  A., Massey, G. (Eds.), Philosophical Problems of the Internal and External
  Worlds: Essays on the Philosophy of Adolph Gr\"unbaum. University of
  Pittsburgh Press, Pittsburgh.

\bibitem[{Stein(1967)}]{SteinNST}
Stein, H., 1967. {Newton}ian space-time. The Texas Quarterly 10, 174--200.

\bibitem[{Wald(1984)}]{Wald}
Wald, R.~M., 1984. General Relativity. University of Chicago Press, Chicago.

\bibitem[{Weatherall(2015)}]{WeatherallTheoreticalEquiv}
Weatherall, J.~O., 2015. Are {Newton}ian gravitation and geometrized
  {Newton}ian gravitation theoretically equivalent? ErkenntnisPublished online.
  doi:10.1007/s10670-015-9783-5.

\bibitem[{Weatherall(2016{\natexlab{a}})}]{WeatherallFBYMGR}
Weatherall, J.~O., 2016{\natexlab{a}}. Fiber bundles, {Yang-Mills} theory, and
  general relativity. Synthese 193~(8), 2389--2425.

\bibitem[{Weatherall(2016{\natexlab{b}})}]{WeatherallSTG}
Weatherall, J.~O., 2016{\natexlab{b}}. Space, time, and geometry from {Newton}
  to {Einstein}, feat. maxwell, lecture notes from the 2016 MCMP summer school
  in mathematical philosophy; available on request.

\bibitem[{Weatherall(2016{\natexlab{c}})}]{WeatherallUG}
Weatherall, J.~O., 2016{\natexlab{c}}. Understanding gauge. Philosophy of
  Science 83~(5), 1039--1049.

\bibitem[{Weatherall(2017)}]{WeatherallCategories}
Weatherall, J.~O., 2017. Categories and the foundations of classical field
  theories. In: Landry, E. (Ed.), Categories for the Working Philosopher.
  Oxford University Press, Oxford, UK, arXiv:1505.07084 [physics.hist-ph].

\bibitem[{Weatherall(2018{\natexlab{a}})}]{WeatherallErratum}
Weatherall, J.~O., 2018{\natexlab{a}}. Erratum. Philosophy of Science 85~(2),
  325.

\bibitem[{Weatherall(2018{\natexlab{b}})}]{WeatherallHoleArg}
Weatherall, J.~O., 2018{\natexlab{b}}. Regarding the ‘hole argument’. The
  British Journal for the Philosophy of Science 69~(2), 329--350.

\bibitem[{Weatherall(2018{\natexlab{c}})}]{WeatherallStein}
Weatherall, J.~O., 2018{\natexlab{c}}. Some philosophical prehistory of the
  ({E}arman-{N}orton) hole argumentArXiv:1812.04574 [physics.hist-ph].

\end{thebibliography}

\end{document}